\documentclass[aps,superscriptaddress,twocolumn,amsfonts,amsmath,showpacs,floatfix]{revtex4}
\usepackage{epsfig}
\usepackage{graphicx}
\input epsf
\epsfclipon

\begin{document}

\title{Phase diagram of silica from computer simulation}

\author{Ivan Saika-Voivod} 
\affiliation{Dipartimento di Fisica and Istituto Nazionale per la
Fisica della Materia, Universita' di Roma La Sapienza, Piazzale Aldo
Moro~2, I-00185, Roma, Italy}

\author{Francesco Sciortino} 
\affiliation{Dipartimento di Fisica and Istituto Nazionale per la
Fisica della Materia, Universita' di Roma La Sapienza, Piazzale Aldo
Moro~2, I-00185, Roma, Italy}
\affiliation{INFM-CRS SOFT:
Complex Dynamics in Structured Systems, Universita' di Roma La
Sapienza, Piazzale Aldo Moro~2, I-00185, Roma, Italy}

\author{Tor Grande} 
\affiliation{Department of Chemistry, Norwegian University of Science 
and Technology, N-7491 Trondheim, Norway}

\author{Peter H. Poole}
\affiliation{Department of Physics, St. Francis Xavier
University, Antigonish, Nova Scotia B2G 2W5, Canada}

\begin{abstract}
We evaluate the phase diagram of the ``BKS'' potential [Van Beest,
Kramer and van Santen, Phys. Rev. Lett. {\bf 64}, 1955 (1990)], a
model of silica widely used in molecular dynamics (MD) simulations. We
conduct MD simulations of the liquid, and three crystals
($\beta$-quartz, coesite and stishovite) over wide ranges of
temperature and density, and evaluate the total Gibbs free energy of
each phase.  The phase boundaries are determined by the intersection
of these free energy surfaces.  Not unexpectedly for a classical pair
potential, our results reveal quantitative discrepancies between the
locations of the BKS and real silica phase boundaries.  At the same
time, we find that the topology of the real phase diagram is
reproduced, confirming that the BKS model provides a satisfactory
qualitative description of a silica-like material.  We also compare
the phase boundaries with the locations of liquid-state thermodynamic
anomalies identified in previous studies of the BKS model.
\end{abstract}

 \maketitle

\section{Introduction}

The melts of silica, water and a number of other molecular substances
(e.g Si, GeO$_2$, BeF$_2$) at ambient pressure $P$ form so-called
``tetrahedral liquids,'' that is, liquids with properties that are
strongly influenced by the occurrence of a network of tetrahedrally
arranged atoms.  Such liquids display a rich spectrum of behavior,
including density maxima~\cite{AK76}, dynamical anomalies~\cite{SL87},
as well as the ability to form numerous crystal
polymorphs~\cite{MSA94}.  Evidence also exists that liquid-liquid
phase transitions occur in some of these
systems~\cite{MS98,SSP01,SA03}.  Yet a detailed understanding of the
commonalities among these materials is hampered by our incomplete
knowledge of their properties under comparable conditions.  For
example, we have extensive knowledge of liquid water for temperatures
$T$ near and above the melting temperature $T_m$ for $T\ge 0.85T_m$,
but the behavior below this range remains a subject of debate~\cite{D03}.
Conversely, we have detailed knowledge of molten silica at ambient $P$
for $T\le T_m$, but a much less complete picture of the behavior at
higher $T$ and $P$~\cite{MSA95}.

Computer simulations have contributed to filling this knowledge gap by
providing numerical estimates of liquid behavior outside the range of
current experiments.  However, a key element has been missing from the
description of many of these model materials: their phase diagrams.
In an experimental study of a molecular liquid, knowledge of the phase
diagram -- that is, the coexistence boundaries demarcating the
stability fields of the liquid, gas and the various crystal phases --
provides a vital reference that elucidates the observed thermodynamic,
dynamic, and structural properties of the liquid phase.  Simulations
of molecular liquids are commonly based on semi-empirical classical
interaction potentials that cannot be expected to precisely reproduce
the experimentally known phase diagrams of the real material.  It is
perhaps for this reason that comprehensive phase diagrams have not yet
been developed for the simulation models used widely to study the
complexities of important molecular liquids, such as water and silica.
However, as a consequence, it has not been possible to
self-consistently relate the behavior found in simulations to the
relevant phase boundaries of the model system, as would normally occur
in an experimental study.

With these motivations, we here focus on the BKS model of
silica~\cite{BKS90}.  The BKS model has played an important role over
the last decade in numerous studies of silica and related materials.
For example, the BKS model has been used in studies of pressure
induced amorphization of quartz~\cite{TK91}, the $\alpha$ to $\beta$
quartz phase transition~\cite{MB01,kimizuka:024105}, the
fragile-to-strong dynamical crossover in liquid
silica~\cite{HK99,SPS01,SSP04}, the possibility of liquid-liquid phase
separation in silica~\cite{SSP01}, and in the study of the generic
topological and entropic properties of random tetrahedral
networks~\cite{vink:245201}.  Despite this interest in the BKS model,
only fragments of specific crystal-crystal phase boundaries have been
located, such as the $\alpha$-$\beta$ quartz transition.  To our
knowledge no data currently exists for the melting lines, though the
liquid-gas coexistence curve has been located for a model similar to
BKS~\cite{GG96}.  In this paper we report the phase diagram of the BKS
model, finding the stability fields in the $P$-$T$ plane for the
liquid phase, and three of the prominent crystal phases of real
silica, stishovite, coesite and $\beta$-quartz.

\section{Methods}

We use the BKS potential, modified at short range to prevent
unphysical ``fusion'' events, and at long range to reduce the system
size dependence of measured properties and to facilitate determination
of minimum energy structures (``inherent structures'').
Ref.~\cite{SSP04} provides a detailed specification of the modified
potential.  The modified potential is indistinguishable from the
original BKS potential in terms of averaged structural and dynamical
behavior.  As described in Ref.~\cite{SSP04}, the values of
thermodynamic properties are slightly shifted compared to the original
BKS potential, but the qualitative behavior is unaffected.  The
Coulombic contribution to the energy is evaluated via the Ewald
method, where the reciprocal space summation is carried out to a
radius of $9$ times the smallest reciprocal cell width~\cite{AT89}.
In all cases, the time step used in our MD simulations is 1~fs.

We restrict our attention to the liquid phase, and the crystal phases
stishovite, coesite and $\beta$-quartz.  A number of other crystal
phases of silica are known.  However, the stability fields of these
three crystals dominate the phase diagram of silica on the widest
scale of $P$ and $T$, and are therefore natural first choices for
examination~\cite{MSA94}.  These three crystals are also
representative of the main types of local coordination structures
found in silica crystals.  The structure of $\beta$-quartz is an open
network of corner-shared SiO$_4$ tetrahedra; coesite is a denser
network of corner-shared SiO$_4$ tetrahedra; and stishovite is a
network of corner and edge shared SiO$_6$ octahedra.  Previous work
has shown that the BKS model is appropriate for studying both low and
high density crystal structures~\cite{TK95}.  To determine the phase
diagram, our approach is to evaluate numerically the Gibbs free energy
$G$ of each of the phases as a function of $P$ and $T$, and then seek
the lines of intersection of these surface functions.

\begin{figure} 
\includegraphics*[width=0.9\linewidth]{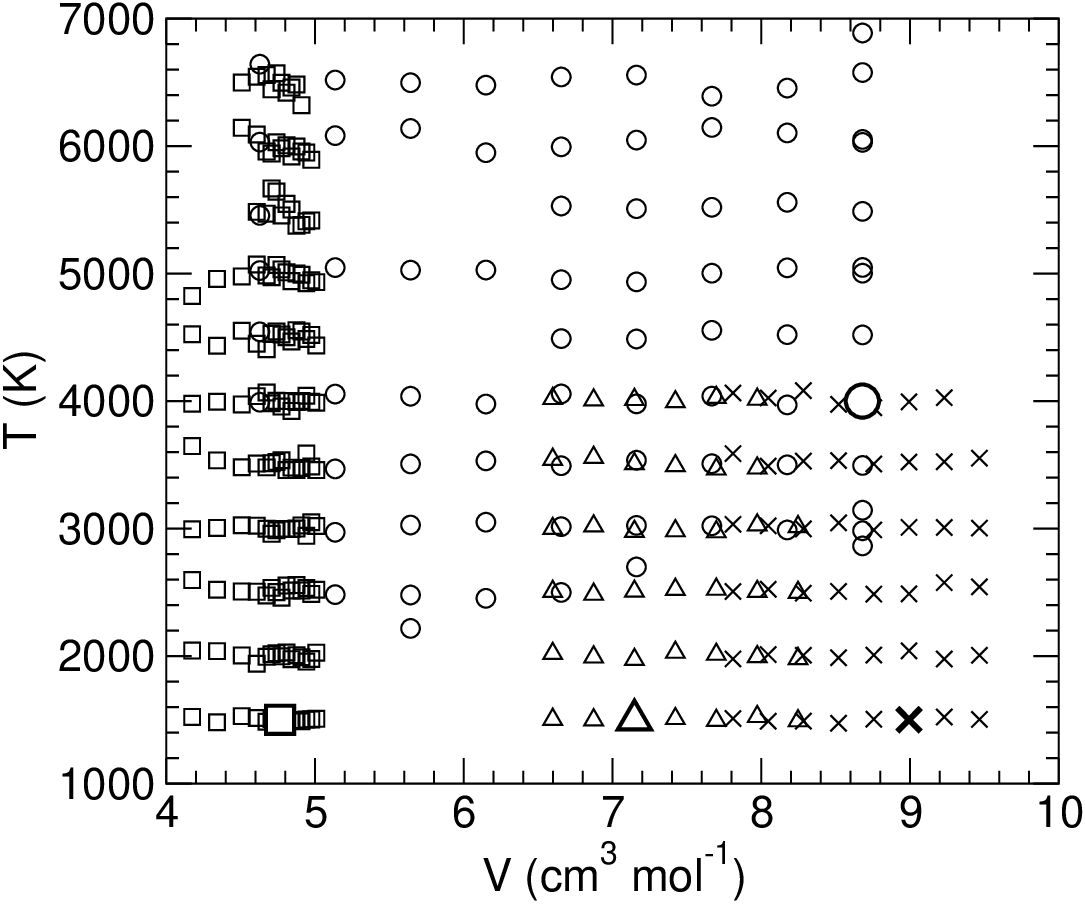}
\caption{\label{map} Location of points in the $V$-$T$ plane at which
we conduct simulations of the liquid (circles), stishovite (squares),
coesite (triangles) and $\beta$-quartz (crosses).  The large symbols
locate reference states $(V_R,T_R)$ at which the entropy of each phase
is evaluated directly.}
\end{figure}

\begin{figure} 
\includegraphics*[width=0.9\linewidth]{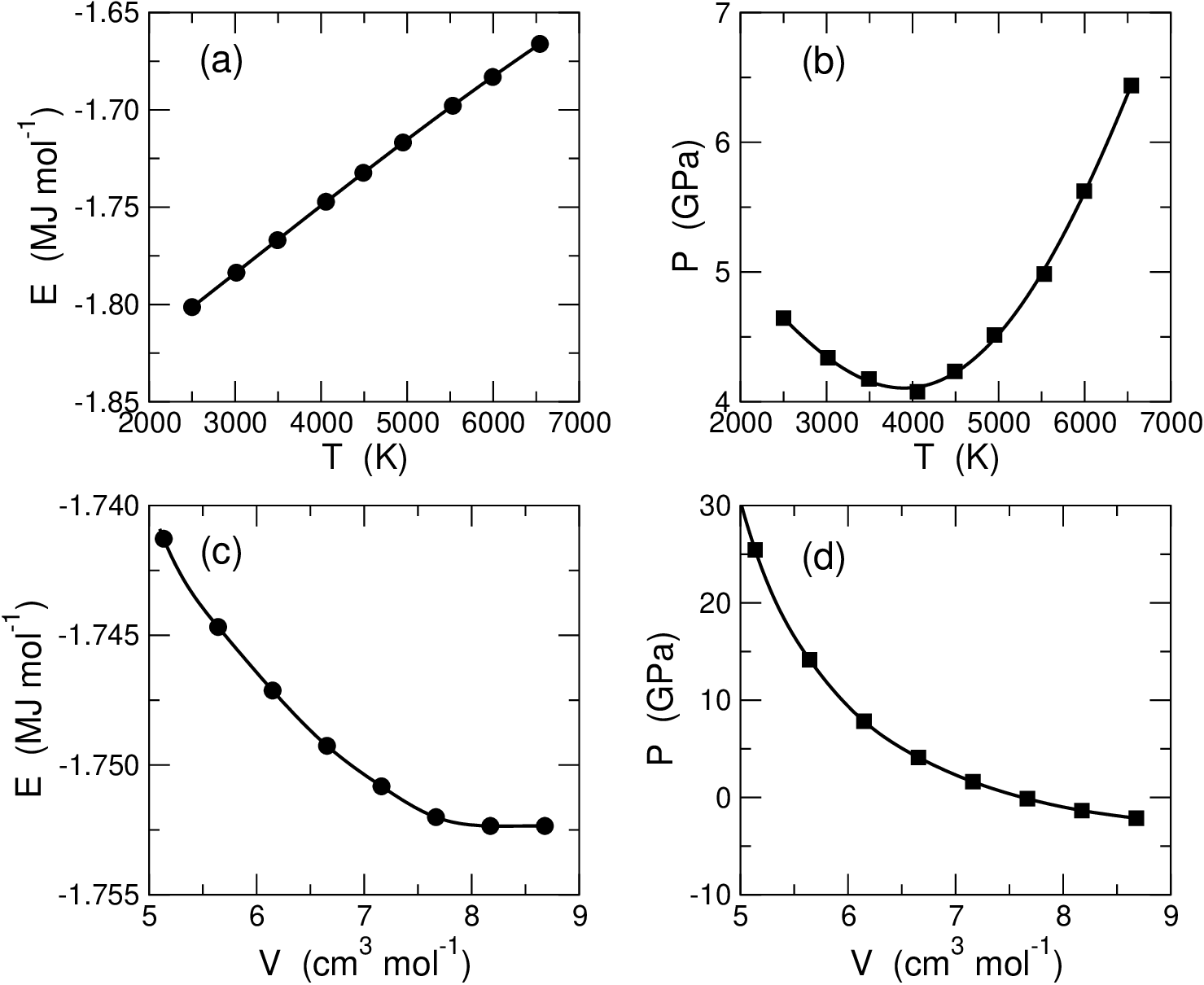}
\caption{\label{fit}Examples of fitted and interpolated data for the
liquid phase: (a) values of $E$ along the $V=6.655$~cm$^3$~mol$^{-1}$
isochore, fitted with a cubic polynomial (line); (b) values of $P$ for
$V=6.655$~cm$^3$~mol$^{-1}$, fitted with a quartic polynomial (line); (c)
interpolated values of $E$ along the $T=4000$~K isotherm, fitted with
a cubic spline (line); (d) interpolated values of $P$ for $T=4000$~K,
fitted with a cubic spline (line).}
\end{figure}

\subsection{Liquid free energy}

For the liquid phase, we use much of the equation of state data
reported in Ref.~\cite{SSP04}, plus some new simulation data generated
using the same methodology.  These simulations modeled a system of a
fixed number of 444 molecular units (1332 ions) in the liquid phase
along nine isochores from volumes $V=4.6296$ to
$8.6804$~cm$^3$~mol$^{-1}$, and ranging in $T$ from nearly 7000~K to
less than 2500~K (Fig.~\ref{map}).  Each of these liquid state points
was equilibrated at constant $V$, and using velocity rescaling to
attain a desired $T$.  Average values of $P$ and $T$ were evaluated
from subsequent constant-$NVE$ runs having a duration of ten times the
time required for silicon atoms to diffuse an average of $0.2$~nm.
The results provide $E(T)$ and $P(T)$ along the specified isochores.
Ref.~\cite{SSP04} also describes the details of a calculation of the
entropy of the liquid phase, $S_R=75.986$ $\pm
0.176$~J~mol$^{-1}$~K$^{-1}$, at a reference state located at
$T_R=4000$~K and $V_R=8.6804$~cm$^3$~mol$^{-1}$.

The value of $E$ at an arbitrary point $(V_o,T_o)$ on the surface
$E(V,T)$ is evaluated as follows.  Along each of the nine isochores
simulated, a 3rd order polynomial in $T$ is fit to the $E$ data.  The
value of $E$ at the desired $T=T_o$ is calculated from the polynomial
found for each $V$.  This creates a set of points approximating the
curve $E(V)$ at $T=T_o$.  A cubic spline passing through these points
is then found, creating a continuous function representating $E(V)$ at
$T=T_o$.  The value of $E$ at the point $(V_o,T_o)$ is evaluated from
this function.  The value of $P$ at an arbitrary point $(V_o,T_o)$ is
calculated in exactly the same way as for $E$, except that a 4th order
polynomial in $T$ is fit to the $P$ data along each of the nine
simulated isochores.  Examples of the simulated and fitted values of
$E$ and $P$ are shown in Fig.~\ref{fit}.

The value of $S$ at an arbitrary point $(V_o,T_o)$ is evaluated by
thermodynamic integration, using the $E(V,T)$ and $P(V,T)$ surfaces
constructed as described above.  The integration is given by,
\begin{equation}\label{TI}
S(V_o,T_o) = S_R
+ \int_{T_R}^{T_o} \frac{1}{T} 
\Biggl(\frac{\partial{E}}{\partial{T}}\Biggr)_{V_R} dT
+ \frac{1}{T_o} \int_{V_R}^{V_o} 
P(V,T_o) dV
\end{equation}
The definite integral over $T$ is evaluated analytically using the
polynomial representation of $E$ as a function of $T$, on the
reference isochore.  The definite integral over $V$ is evaluated
numerically via Simpson's rule, using data from the cubic spline
representation of $P$ as a function of $V$, constructed along the
desired isotherm.

We combine these numerical estimates to determine the Gibbs free
energy $G$ at arbitrary state points using
$G(V,T)=E(V,T)+VP(V,T)-TS(V,T)$.  To find $G$ at an arbitrary $(P,T)$
point, we find the value of $V$ from $P(V,T)$ such that $P$ has the
desired value.  In this way we can construct arbitrary isotherms or
isobars cutting through the $G(P,T)$ surface.

\begin{figure} 
\includegraphics*[width=0.8\linewidth]{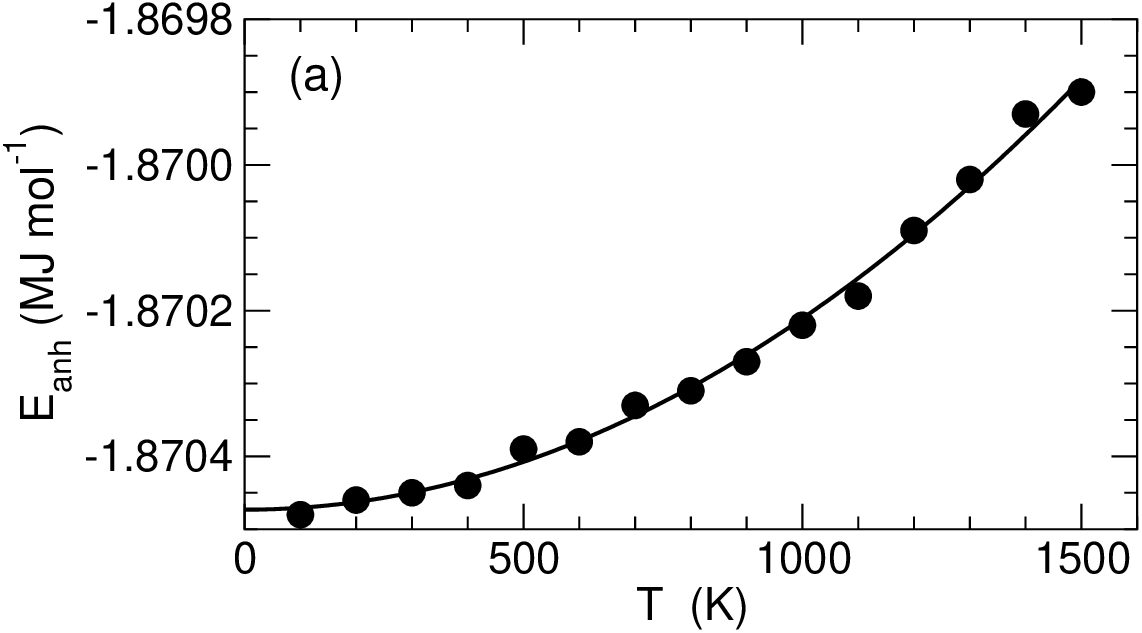}
\includegraphics*[width=0.8\linewidth]{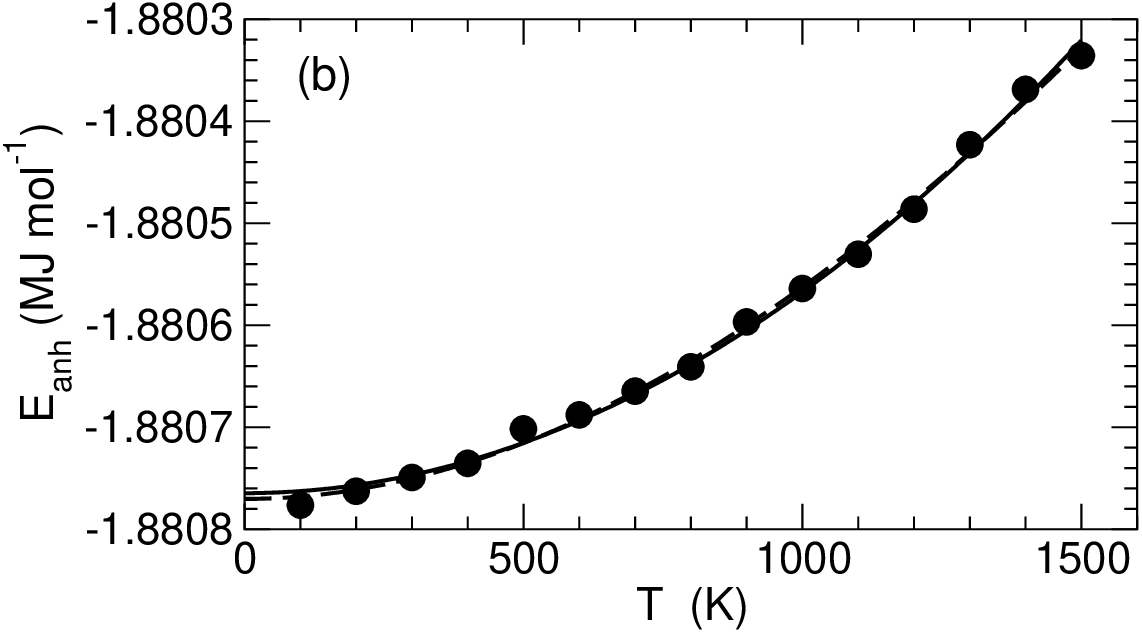}
\includegraphics*[width=0.8\linewidth]{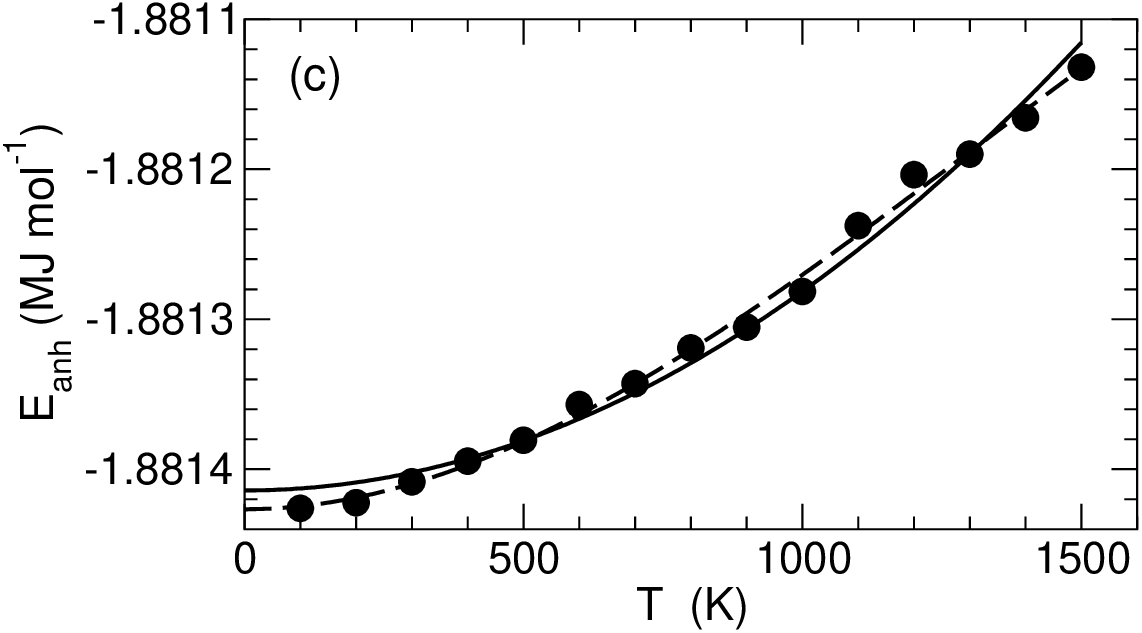}
\caption{\label{anh} $E_{anh}(T)$ for (a) $\beta$-quartz, (b) coesite
and (c) stishovite along their respective reference isochores.  In (a)
the solid line is the fit to the data given by
Eq.~\protect{\ref{efit}} with $N_{max}=2$. The fit for $N_{max}=3$ is
not visible as it overlaps with the $N_{max}=2$ curve on the scale of
this plot.  In (b) both the fit for $N_{max}=2$ (solid) and
$N_{max}=3$ (dashed) are shown.  In (c) the fits for $N_{max}=2$
(solid) and $N_{max}=4$ (dashed) are shown, while the curve for
$N_{max}=3$ is not visible as it overlaps with the $N_{max}=4$ curve.}
\end{figure}

\subsection{Crystal free energy}

We conduct simulations of three crystal phases: stishovite,
$\beta$-quartz and coesite.  Our simulations employ 1200 ions for
stishovite, 1536 ions for coesite, and 1350 ions for $\beta$-quartz.
We carry out simulations over a range of $V$ and $T$ appropriate for
each phase, as shown in Fig.~\ref{map}.

We employ the following procedure to obtain equilibrium averages for
$E(T)$ and $P(T)$ along the specified isochores.  The rationale
underlying this procedure is to allow us to obtain thermodynamic
properties along a set of specified isochores, so that we may
construct the $G(P,T)$ surface for each phase in the same way as
described above for the liquid phase. However, obtaining isochoric
data for crystals requires care, as unit cell parameters may change
with $T$, even though the overall density remains fixed.  In
particular, we must ensure that anisotropic stresses do not arise in
the simulation cell.  The procedure, for each crystal, is as follows:

First, we create an initial configuration of stishovite,
$\beta$-quartz~\cite{KC92} and coesite~\cite{SSAK87}.  Then for a
number of specified $V$, we optimize the $T=0$ atomic coordinates and
unit cell parameters to minimize the energy and to remove anisotropic
stresses.  This optimization is carried out at constant overall $V$,
and consists of alternating applications of the simplex method (to
optimize cell parameters) and the conjugate gradient method (to
optimize atomic coordinates)~\cite{NR92}.  This optimization cycle is
repeated until the energy converges to a minimum value to within a
tolerance of $10^{-10}$.

Then, for each $T$ at which we desire thermodynamic properties, we
carry out the following steps:

(i) Beginning with the optimized configuration at the appropriate $V$,
we conduct a 20~ps constant $V$ simulation, during which the desired
$T$ is established via velocity rescaling every $100$ time steps.

(ii) The configuration produced in (i) is used to initiate a $20$~ps
constant-$NVE$ simulation, to ensure that an equilibrium state at the
desired $T$ has been achieved.

(iii) To relax any anisotropic stress that may have arisen in bringing
the system to non-zero $T$, we carry out a 40~ps constant-$NPT$
simulation (during which the simulation cell geometry is
unconstrained) where we set $P$ to the average value from step (ii).

(iv) Step (iii) may have changed the overall $V$ of the simulation
cell away from the desired isochore.  We restore the value of $V$ of
the simulation cell by isotropically rescaling the average cell
lengths obtained in step (iii), while leaving the obtained average
angles fixed.  For all crystals, the rescaling is never more than
0.5\% of the desired volume, and is typically 0.1\%. This rescaled
configuration is then used to initiate a 30~ps constant-$NVE$
simulation, during which the average values of $P$ and $T$ are
evaluated.

We note that for $\beta$-quartz, step~(iii) is carried out for $50$~ps
and step~(iv) for $80$~ps.  These longer times are used in order to
resolve the subtle variation of $P$ along isochores, since
$\beta$-quartz displays a density maximum in the region of our
simulations.  We also note that the $\beta$-quartz phase spontaneously
converts to $\alpha$-quartz, but only for $T$ and $V$ outside the
range of simulated points shown for $\beta$-quartz in Fig.~\ref{map}.
Our results therefore pertain only to $\beta$-quartz and are not
influenced by this crystal-crystal phase transition.

The above procedure provides $E(T)$ and $P(T)$ along specified
isochores.  Using the same fitting and interpolation procedure as is
used for the liquid, we can therefore evaluate $E$ and $P$ at
arbitrary state points $(V,T)$.

Finally, we need to evaluate $S$ for each crystal at a reference state
point, in order to construct the surface $S(V,T)$ via thermodynamic
integration.  Our method is as follows.  For each crystal phase we
select a reference volume $V_R$ (see Fig.~\ref{map}), and choose the
$T_R=1500$~K configuration obtained at the end of step~(iv) above.
Using the conjugate gradient method, we optimize the atomic positions
(at fixed cell geometry) to find the minimum energy configuration.  We
then evaluate the Hessian matrix of this minimum energy configuration
and diagonalize it to find the eigenfrequency spectrum.  The classical
harmonic entropy is found from this eigenfrequency spectrum.  (The
details of this approach are given in Ref.~\cite{SSP04}, where the
method is used to find the classical harmonic entropy of inherent
structures of the liquid state.)

To determine the total entropy, we need to evaluate the anharmonic
contribution and add it to the harmonic entropy found above.  We use
the energy-optimized configuration for which we calculate the harmonic
entropy as the starting configuration for $15$ equally spaced
simulations from $T=100$~K to $T_R=1500$~K.  We simulate each state
point at constant $V$ using velocity scaling to maintain $T$ at the
desired value, with fixed cell geometry, for $150$~ps ($400$~ps for
stishovite).  From these simulations we evaluate,
\begin{equation}
E_{anh}(T) = U(T) - \frac{3}{2} R (1 - 1/N)T,
\end{equation}
where $U$ is the potential energy and $R$ is the gas constant.  Using
a polynomial fit,
\begin{equation}
E_{anh} = a_0 + \sum_{n=2}^{N_{max}} a_n T^n, 
\label{efit}
\end{equation}
we evaluate, 
\begin{equation}
S_{anh}(T_R) =
\int_0^{T_R} \frac{1}{T} 
\Biggl (\frac{\partial E_{anh}(T)}{\partial T}\Biggr)_V dT.
\end{equation}
By using different values of $N_{max}=2$ and $3$ for coesite and
$\beta$-quartz, and $3$ and $4$ for stishovite, we obtain error
estimates for $S_{anh}(T_R)$.  Fig.~\ref{anh} shows the variation of
$E_{anh}$ with $T$ for each of the three crystals simulated.  The
resulting reference entropies for each crystal phase at $T_R=1500$~K
at their respective reference volumes $V_R$ are as follows:
44.982~J~mol$^{-1}$~K$^{-1}$ at $V_R=8.9933$~cm$^3$~mol$^{-1}$ for
$\beta$-quartz; 43.682~J~mol$^{-1}$~K$^{-1}$ at
$V_R=7.1478$~cm$^3$~mol$^{-1}$ for coesite; and
39.536~J~mol$^{-1}$~K$^{-1}$ at $V_R=4.7650$~cm$^3$~mol$^{-1}$ for
stishovite.  The uncertainty in each $S$ value is approximately
$0.01$~J~mol$^{-1}~$K$^{-1}$.

The above procedure provides $E(V,T)$, $P(V,T)$ and the reference
value of $S$ for each of the three crystal phases.  The procedure to
evaluate $G(P,T)$ from this information for each crystal phase is the
same as is used for the liquid phase.

\begin{figure} 
\includegraphics*[width=0.9\linewidth]{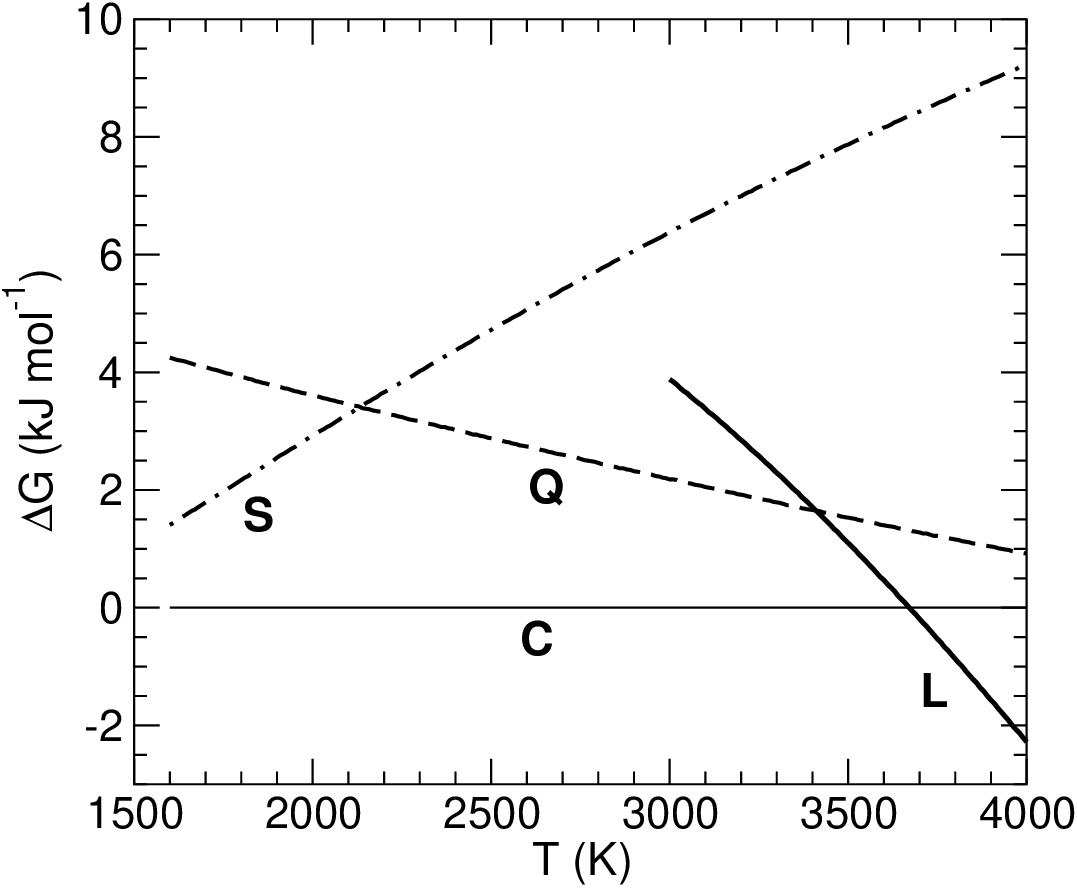}
\caption{\label{dg} $\Delta G$, the Gibbs free energy relative to that
of coesite (C) for the liquid (L), $\beta$-quartz (Q) and stishovite
(S) phases, at constant $P=2$~GPa.  The intersections locate points on
the stable and metastable coexistence lines that cross this isobar.}
\end{figure}

\subsection{Coexistence boundaries and error estimates}

For every pair of phases we determine the coexistence line as the
locus of points in the plane of $P$ and $T$ for which $G$ for the two
phases is the same.  Along each locus, we also find the value of $V$
for each of the two coexisting phases.  Fig.~\ref{dg} shows an example
of the intersection of isobars of $G$ (relative to coesite) for each
phase at $P=2$~GPa.

We perform several checks on our scheme.  We calculate the change in
$S$ for a single phase around a closed path in the $V$-$T$ plane,
which we find to be zero within an error of approximately
$0.01$~J~mol$^{-1}$~K$^{-1}$.  We also check that the relations
$P=-(\partial A/\partial V)_T$ (where $A$ is the Helmholtz free
energy) and $P-T(\partial P/\partial T)_V = - (\partial E/\partial
V)_T$ are satisfactorily met.  Furthermore, along the coexistence
lines, we check the Clapeyron relation $dP/dT=\Delta S/\Delta V$,
where $\Delta S$ is the difference in $S$ between the two phases and
$\Delta V$ is the difference in $V$; we find this to be satisfied to
within $0.1$~MPa K$^{-1}$.

\begin{figure}
\includegraphics*[width=0.9\linewidth]{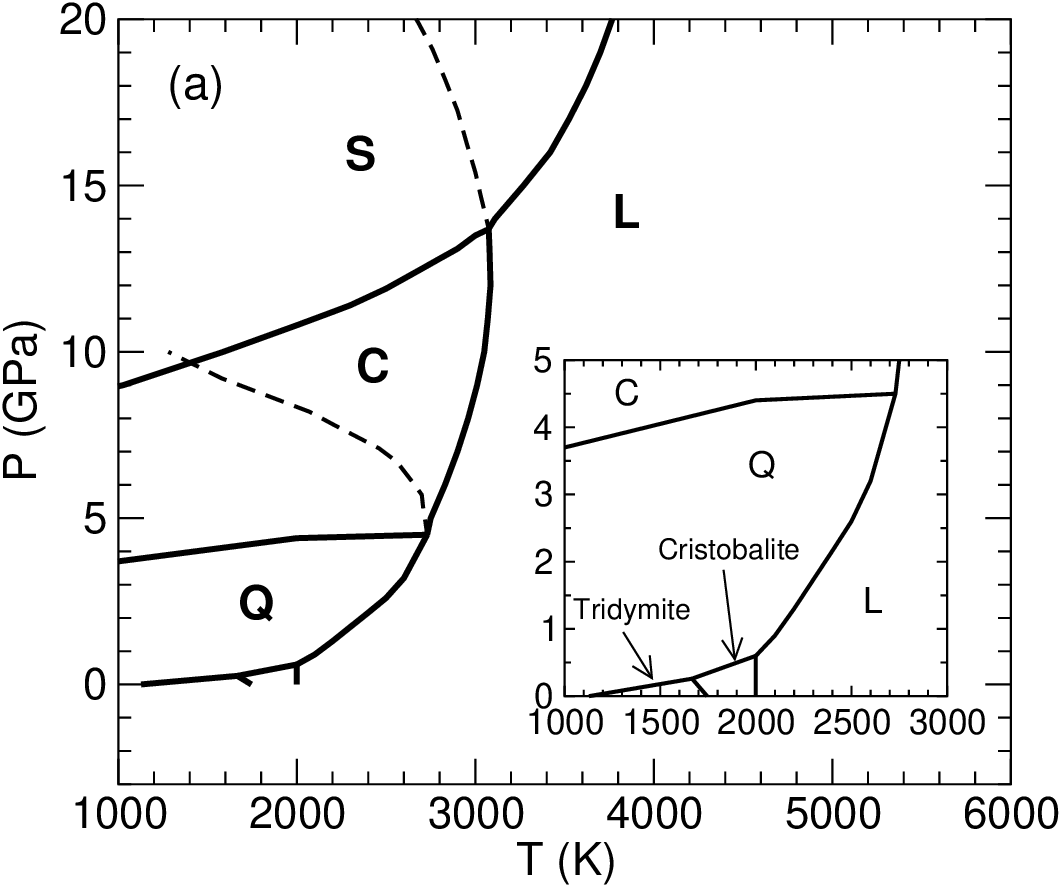}
\includegraphics*[width=0.9\linewidth]{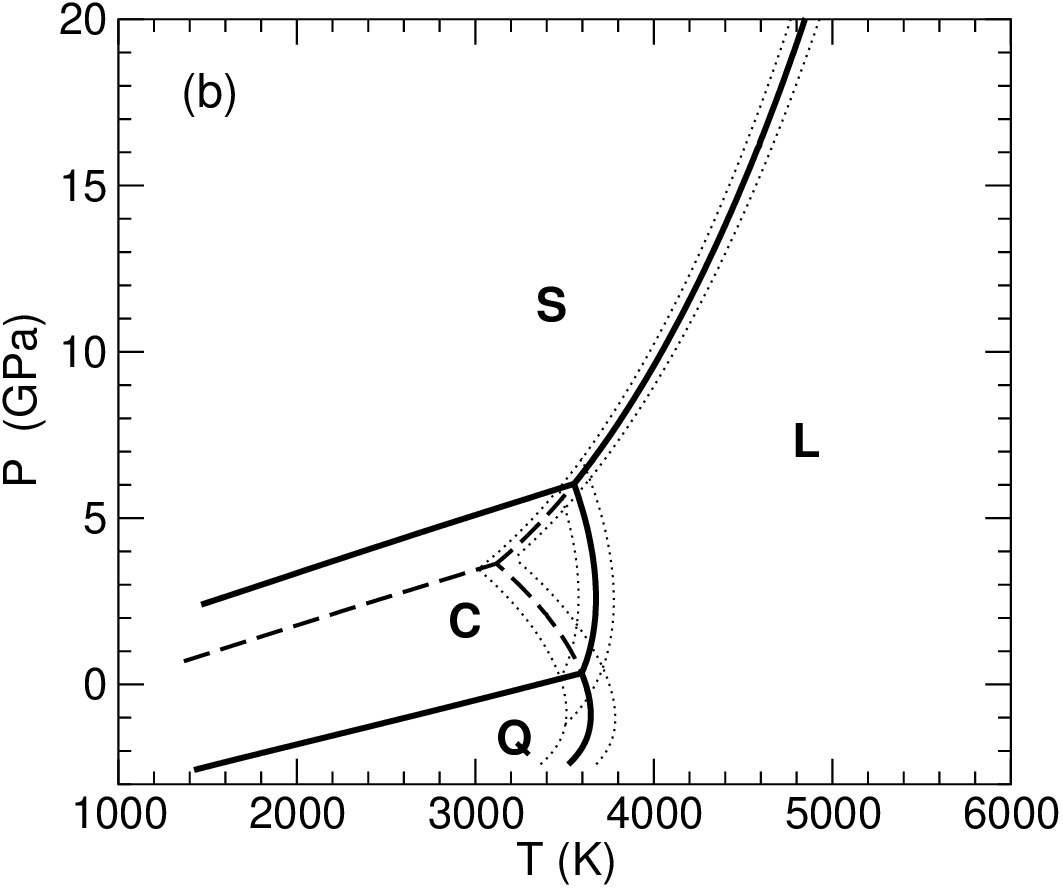}
\caption{\label{PD} (a) Experimentally determined coexistence lines of
silica in the $P$-$T$ plane.  Stability fields for the stishovite (S),
coesite (C), $\beta$-quartz (Q) and liquid (L) phases are shown.  Both
stable (solid) and metastable (dashed) coexistence lines are shown.
The inset shows the stability fields of cristobalite and tridymite,
not considered in this work.  Adapted from
Ref.~\protect{\cite{MSA94}}.  (b) Phase diagram of BKS silica in the
$P$-$T$ plane.  Solid lines are stable coexistence lines.  Dotted
lines show error estimates for the crystal-liquid coexistence lines,
as described in the text.  Metastable coexistence lines (dashed) are
also shown that meet at the metastable S-L-Q triple point.}
\end{figure}

Throughout the evaluation scheme described above, the largest single
source of statistical error is the uncertainty cited in
Ref.~\cite{SSP04} for $S_R$, the entropy of the liquid at the
reference state point.  We therefore create confidence limits for our
melting lines, shown in Fig.~\ref{PD}, by allowing the value of $S_R$
to vary by $\pm 0.18$~J~mol$^{-1}$~K$^{-1}$.

\begin{figure} 
\includegraphics*[width=0.9\linewidth]{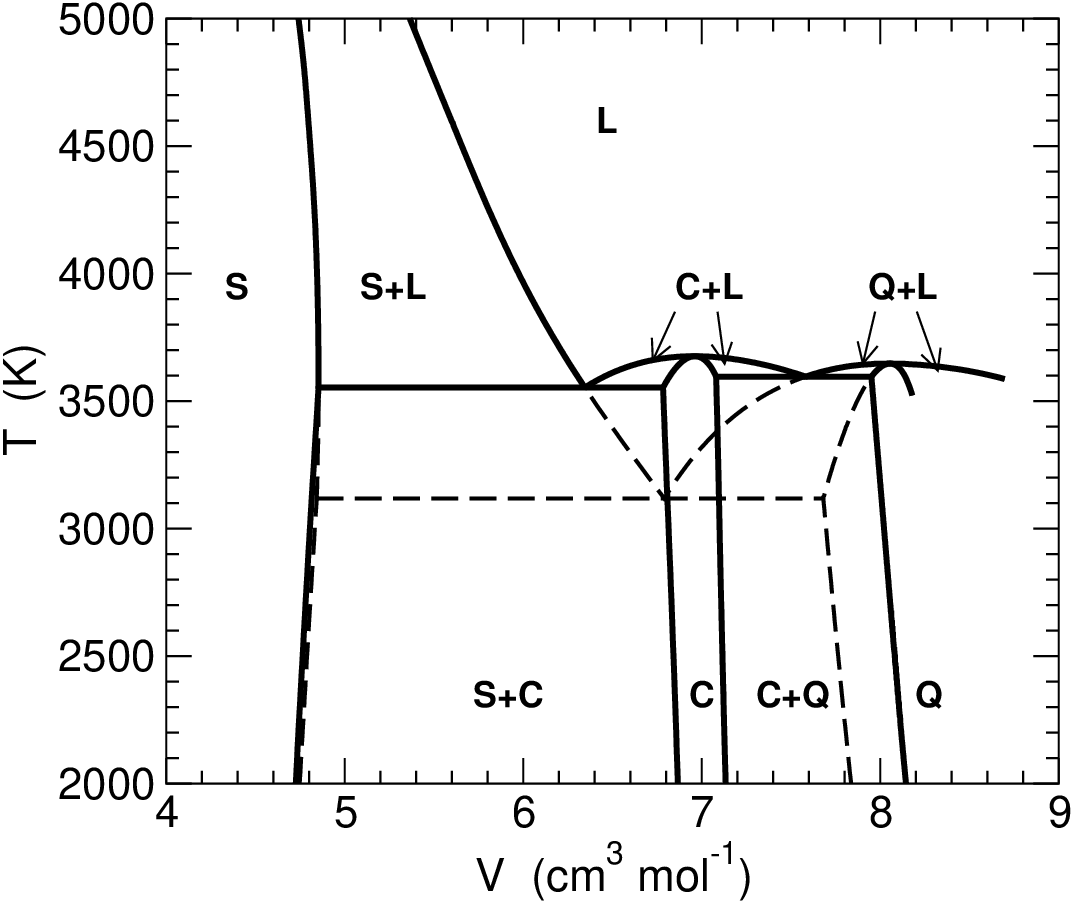}
\caption{\label{VD} Phase diagram of BKS silica in the $V$-$T$ plane.
The notation and symbols used have the same meaning as in
Fig.~\protect{\ref{PD}}.  Note that in this projection, both one-phase
stability fields as well as two-phase coexistence regions are located.
The projections of the metastable coexistence lines (dashed) shown in
Fig.~\protect{\ref{PD}} are also presented.}
\end{figure}

\begin{figure} 
\includegraphics*[width=0.9\linewidth]{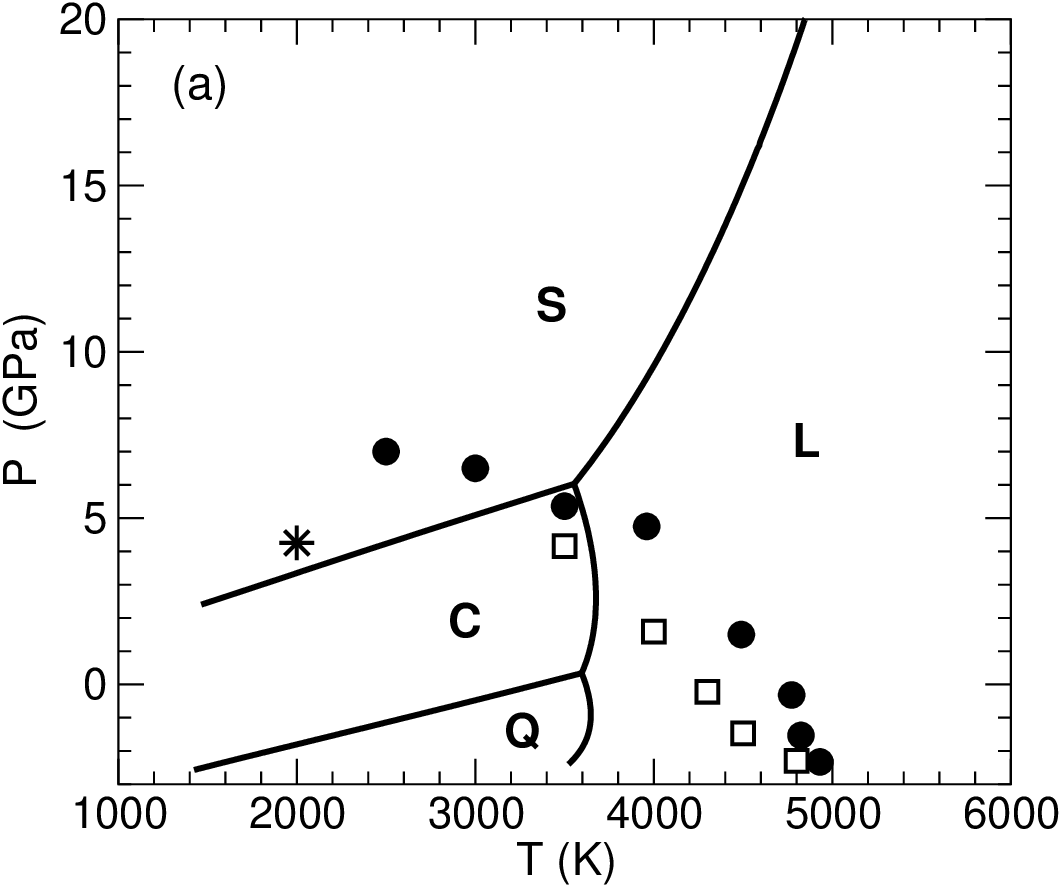}
\includegraphics*[width=0.9\linewidth]{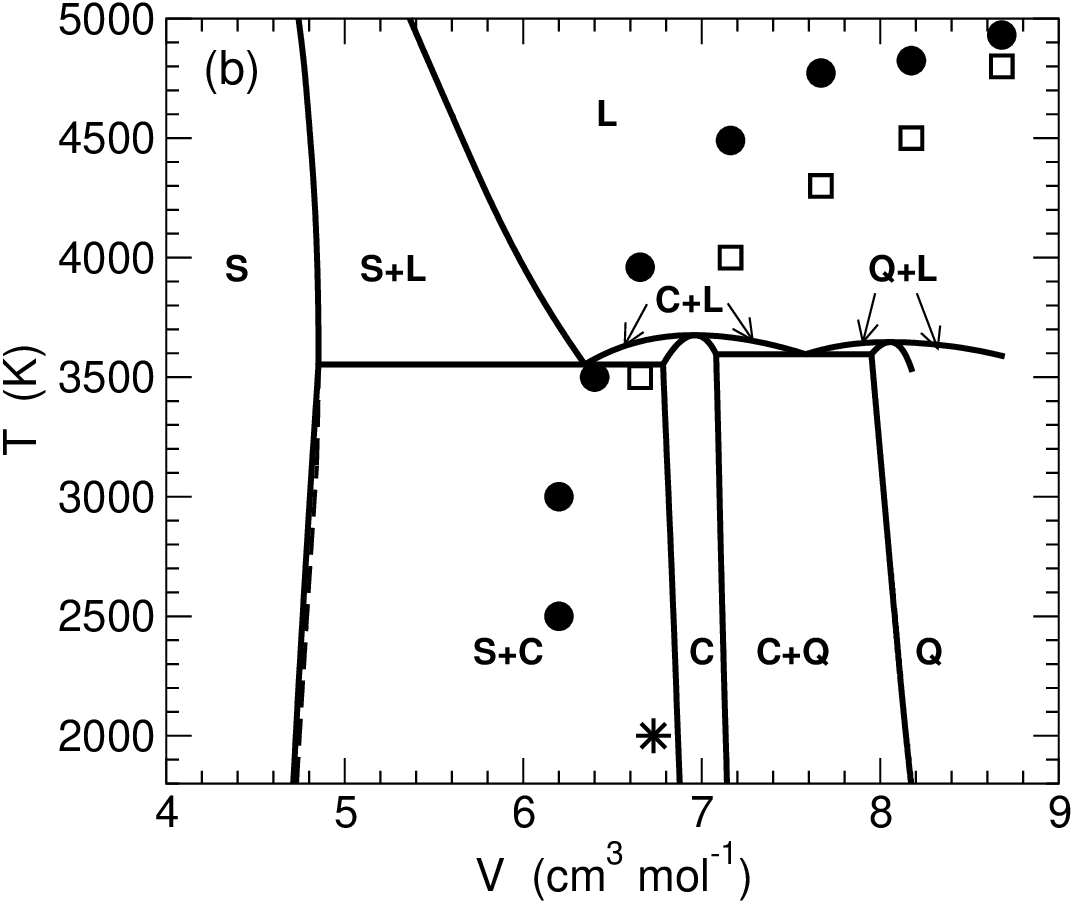}
\caption{\label{aug} BKS phase boundaries in (a) the $P$-$T$ plane and
(b) the $V$-$T$ plane, in relation to density maxima (filled circles)
and $C_V$ maxima (squares) in the liquid phase.  Also located is the
state point (star) at which evidence of liquid-liquid phase separation
was reported in Ref.~\protect{\cite{SSP01}}.}
\end{figure}

\section{Results and Discussion}

Fig.~\ref{PD}(b) plots $P$-$T$ coexistence conditions, both stable and
metastable, occurring among the liquid phase (L) and the crystalline
phases $\beta$-quartz (Q), coesite (C) and stishovite (S).
Fig.~\ref{VD} is the projection of the same boundaries onto the plane
of $V$ and $T$.  This plot exposes the volume differences of
coexisting phases along phase boundaries.  This type of plot is rarely
constructed for real materials, due to the challenge of determining
the densities of coexisting phases, especially at high pressure.
However, it is readily constructed from simulation data.

Comparison of the BKS and experimental phase boundaries~\cite{MSA94}
in Fig.~\ref{PD} exposes the quantitative deficiencies of the model.
Apparent in particular is the difference between the pressures at
which corresponding features occur.  For example, the S-L-C triple
point occurs at 13.4~GPa in real silica, but at only 5.8~GPa in the
model.  Overall, the $P$ range of the crystal stability fields is
substantially lower in the model.  The correspondence of the thermal
behavior is better, but significant differences still occur.  The $T$
of the S-L-C and C-L-Q triple points are respectively 15\% and 32\%
higher than their experimental values.

However, it is noteworthy that the topology of the real silica phase
diagram is reproduced by the model.  All three studied crystals have
large stability fields, which increase in extent as $P$ increases.
More subtle features, notably the melting line maxima in both the Q-L
and C-L coexistence lines are also reproduced.  Also, the occurrence
of a metastable S-Q-L triple point in the stability field of coesite,
suggested by an extrapolation of the experimental boundaries, is
observed in the model.  Thus, despite its quantitative deficiencies,
the BKS model is appropriate for studying the qualitative behavior of
a substance with a silica-like phase diagram.

The phase information given in Figs.~\ref{PD} and \ref{VD} allows
previous (and future) observations of the behavior of BKS silica to be
considered within the context of the phase behavior of the model
itself.  For example, several studies of BKS silica have identified
the location of a density maximum in the liquid
phase~\cite{HK99,SSP01}.  A thermal anomaly, a line of maxima of the
isochoric specific heat $C_V$, has also been located in simulations of
the liquid~\cite{SPS01,SSP04}.  Both of these features have been
related to the early stages of the formation of a structured
tetrahedral network in the liquid state.  This structural evolution
also is believed to underlie a crossover from non-Arrhenian
(``fragile'') to Arrhenian (``strong'') dynamics in the
liquid~\cite{HK99,SPS01,SSP04}.

We show in Fig.~\ref{aug} the location of the line of density and
$C_V$ maxima in both the $P$-$T$ and $V$-$T$ planes.  These lines
approximately separate the liquid behavior into a ``tetrahedral network
influenced'' region at low $T$ and large $V$ (low $P$), and a ``normal
liquid'' region at high $T$ and small $V$ (high $P$).  Consistent with
this, the stability fields of $\beta$-quartz and coesite (both of
which have four-coordinated silicon atoms) occur within the network
influenced region, while the stability field of stishovite (with
six-coordinated silicons) falls outside.  

We also show in Fig.~\ref{aug} the location of the state point at
which evidence of liquid-liquid phase separation was reported in
Ref.~\cite{SSP01}.  This point occurs at a density just above that of
the high density edge of the one-phase stability field of coesite.
This is a plausible density at which the open tetrahedral network
structure of the (one-phase) supercooled liquid state begins to
collapse to a higher density, perhaps via a discontinuous phase
transition.

\begin{acknowledgements}
This research was supported by NSERC (Canada), SHARCNET, the Canada
Research Chairs Program, MIUR FIRB and PRIN 2000.
\end{acknowledgements}

\bibliography{refs}

\begin{thebibliography}{22}
\expandafter\ifx\csname natexlab\endcsname\relax\def\natexlab#1{#1}\fi
\expandafter\ifx\csname bibnamefont\endcsname\relax
  \def\bibnamefont#1{#1}\fi
\expandafter\ifx\csname bibfnamefont\endcsname\relax
  \def\bibfnamefont#1{#1}\fi
\expandafter\ifx\csname citenamefont\endcsname\relax
  \def\citenamefont#1{#1}\fi
\expandafter\ifx\csname url\endcsname\relax
  \def\url#1{\texttt{#1}}\fi
\expandafter\ifx\csname urlprefix\endcsname\relax\def\urlprefix{URL }\fi
\providecommand{\bibinfo}[2]{#2}
\providecommand{\eprint}[2][]{\url{#2}}

\bibitem[{\citenamefont{Angell and Kanno}(1976)}]{AK76}
\bibinfo{author}{\bibfnamefont{C.}~\bibnamefont{Angell}} \bibnamefont{and}
  \bibinfo{author}{\bibfnamefont{H.}~\bibnamefont{Kanno}},
  \bibinfo{journal}{Science} \textbf{\bibinfo{volume}{193}},
  \bibinfo{pages}{1121} (\bibinfo{year}{1976}).

\bibitem[{\citenamefont{Prielmeier et~al.}(1987)\citenamefont{Prielmeier, Lang,
  Speedy, and Ludemann}}]{SL87}
\bibinfo{author}{\bibfnamefont{F.}~\bibnamefont{Prielmeier}},
  \bibinfo{author}{\bibfnamefont{E.}~\bibnamefont{Lang}},
  \bibinfo{author}{\bibfnamefont{R.}~\bibnamefont{Speedy}}, \bibnamefont{and}
  \bibinfo{author}{\bibfnamefont{H.}~\bibnamefont{Ludemann}},
  \bibinfo{journal}{Phys. Rev. Lett.} \textbf{\bibinfo{volume}{59}},
  \bibinfo{pages}{1128} (\bibinfo{year}{1987}).

\bibitem[{\citenamefont{Heaney et~al.}(1994)\citenamefont{Heaney, Prewitt, and
  Gibbs}}]{MSA94}
\bibinfo{editor}{\bibfnamefont{P.}~\bibnamefont{Heaney}},
  \bibinfo{editor}{\bibfnamefont{C.}~\bibnamefont{Prewitt}}, \bibnamefont{and}
  \bibinfo{editor}{\bibfnamefont{G.}~\bibnamefont{Gibbs}}, eds.,
  \emph{\bibinfo{title}{Silica: physical behavior, geochemistry and materials
  applications}}, vol.~\bibinfo{volume}{29} of \emph{\bibinfo{series}{Reviews
  in Mineralogy}} (\bibinfo{publisher}{Mineralogical Society of America},
  \bibinfo{year}{1994}).

\bibitem[{\citenamefont{Mishima and Stanley}(1998)}]{MS98}
\bibinfo{author}{\bibfnamefont{O.}~\bibnamefont{Mishima}} \bibnamefont{and}
  \bibinfo{author}{\bibfnamefont{H.~E.} \bibnamefont{Stanley}},
  \bibinfo{journal}{Nature} \textbf{\bibinfo{volume}{396}},
  \bibinfo{pages}{329} (\bibinfo{year}{1998}).

\bibitem[{\citenamefont{Saika-Voivod
  et~al.}(2001{\natexlab{a}})\citenamefont{Saika-Voivod, Sciortino, and
  Poole}}]{SSP01}
\bibinfo{author}{\bibfnamefont{I.}~\bibnamefont{Saika-Voivod}},
  \bibinfo{author}{\bibfnamefont{F.}~\bibnamefont{Sciortino}},
  \bibnamefont{and} \bibinfo{author}{\bibfnamefont{P.~H.} \bibnamefont{Poole}},
  \bibinfo{journal}{Phys. Rev. E} \textbf{\bibinfo{volume}{63}},
  \bibinfo{pages}{011202} (\bibinfo{year}{2001}{\natexlab{a}}).

\bibitem[{\citenamefont{Sastry and Angell}(2003)}]{SA03}
\bibinfo{author}{\bibfnamefont{S.}~\bibnamefont{Sastry}} \bibnamefont{and}
  \bibinfo{author}{\bibfnamefont{C.~A.} \bibnamefont{Angell}},
  \bibinfo{journal}{Nature Materials} \textbf{\bibinfo{volume}{2}},
  \bibinfo{pages}{739} (\bibinfo{year}{2003}).

\bibitem[{\citenamefont{Debenedetti}(2003)}]{D03}
\bibinfo{author}{\bibfnamefont{P.~G.} \bibnamefont{Debenedetti}},
  \bibinfo{journal}{J. Phys.: Condens. Matter} \textbf{\bibinfo{volume}{15}},
  \bibinfo{pages}{R1669} (\bibinfo{year}{2003}).

\bibitem[{\citenamefont{Stebbins et~al.}(1995)\citenamefont{Stebbins, McMillan,
  and Dingwell}}]{MSA95}
\bibinfo{editor}{\bibfnamefont{J.~F.} \bibnamefont{Stebbins}},
  \bibinfo{editor}{\bibfnamefont{P.}~\bibnamefont{McMillan}}, \bibnamefont{and}
  \bibinfo{editor}{\bibfnamefont{D.}~\bibnamefont{Dingwell}}, eds.,
  \emph{\bibinfo{title}{Structure, dynamics and properties of silicate melts}},
  vol.~\bibinfo{volume}{32} of \emph{\bibinfo{series}{Reviews in Mineralogy}}
  (\bibinfo{publisher}{Mineralogical Society of America},
  \bibinfo{year}{1995}).

\bibitem[{\citenamefont{van Beest et~al.}(1990)\citenamefont{van Beest, Kramer,
  and van Santen}}]{BKS90}
\bibinfo{author}{\bibfnamefont{B.}~\bibnamefont{van Beest}},
  \bibinfo{author}{\bibfnamefont{G.}~\bibnamefont{Kramer}}, \bibnamefont{and}
  \bibinfo{author}{\bibfnamefont{R.}~\bibnamefont{van Santen}},
  \bibinfo{journal}{Phys. Rev. Lett.} \textbf{\bibinfo{volume}{64}},
  \bibinfo{pages}{1955} (\bibinfo{year}{1990}).

\bibitem[{\citenamefont{Tse and Klug}(1991)}]{TK91}
\bibinfo{author}{\bibfnamefont{J.~S.} \bibnamefont{Tse}} \bibnamefont{and}
  \bibinfo{author}{\bibfnamefont{D.~D.} \bibnamefont{Klug}},
  \bibinfo{journal}{Phys. Rev. Lett.} \textbf{\bibinfo{volume}{67}},
  \bibinfo{pages}{3559} (\bibinfo{year}{1991}).

\bibitem[{\citenamefont{Kimizuka et~al.}(2003)\citenamefont{Kimizuka, Kaburaki,
  and Kogure}}]{kimizuka:024105}
\bibinfo{author}{\bibfnamefont{H.}~\bibnamefont{Kimizuka}},
  \bibinfo{author}{\bibfnamefont{H.}~\bibnamefont{Kaburaki}}, \bibnamefont{and}
  \bibinfo{author}{\bibfnamefont{Y.}~\bibnamefont{Kogure}},
  \bibinfo{journal}{Phys. Rev. B} \textbf{\bibinfo{volume}{67}},
  \bibinfo{eid}{024105} (\bibinfo{year}{2003}).

\bibitem[{\citenamefont{Muser and Binder}(2001)}]{MB01}
\bibinfo{author}{\bibfnamefont{M.}~\bibnamefont{Muser}} \bibnamefont{and}
  \bibinfo{author}{\bibfnamefont{K.}~\bibnamefont{Binder}},
  \bibinfo{journal}{Phys. Chem. Minerals} \textbf{\bibinfo{volume}{28}},
  \bibinfo{pages}{746} (\bibinfo{year}{2001}).

\bibitem[{\citenamefont{Horbach and Kob}(1999)}]{HK99}
\bibinfo{author}{\bibfnamefont{J.}~\bibnamefont{Horbach}} \bibnamefont{and}
  \bibinfo{author}{\bibfnamefont{W.}~\bibnamefont{Kob}},
  \bibinfo{journal}{Phys. Rev. B} \textbf{\bibinfo{volume}{60}},
  \bibinfo{pages}{3169} (\bibinfo{year}{1999}).

\bibitem[{\citenamefont{Saika-Voivod
  et~al.}(2001{\natexlab{b}})\citenamefont{Saika-Voivod, Poole, and
  Sciortino}}]{SPS01}
\bibinfo{author}{\bibfnamefont{I.}~\bibnamefont{Saika-Voivod}},
  \bibinfo{author}{\bibfnamefont{P.~H.} \bibnamefont{Poole}}, \bibnamefont{and}
  \bibinfo{author}{\bibfnamefont{F.}~\bibnamefont{Sciortino}},
  \bibinfo{journal}{Nature} \textbf{\bibinfo{volume}{412}},
  \bibinfo{pages}{514} (\bibinfo{year}{2001}{\natexlab{b}}).

\bibitem[{\citenamefont{Saika-Voivod et~al.}(2004)\citenamefont{Saika-Voivod,
  Sciortino, and Poole}}]{SSP04}
\bibinfo{author}{\bibfnamefont{I.}~\bibnamefont{Saika-Voivod}},
  \bibinfo{author}{\bibfnamefont{F.}~\bibnamefont{Sciortino}},
  \bibnamefont{and} \bibinfo{author}{\bibfnamefont{P.~H.} \bibnamefont{Poole}},
  \bibinfo{journal}{Phys. Rev. E} \textbf{\bibinfo{volume}{69}},
  \bibinfo{eid}{041503} (\bibinfo{year}{2004}).

\bibitem[{\citenamefont{Vink and Barkema}(2003)}]{vink:245201}
\bibinfo{author}{\bibfnamefont{R.}~\bibnamefont{Vink}} \bibnamefont{and}
  \bibinfo{author}{\bibfnamefont{G.}~\bibnamefont{Barkema}},
  \bibinfo{journal}{Phys. Rev. B} \textbf{\bibinfo{volume}{67}},
  \bibinfo{eid}{245201} (\bibinfo{year}{2003}).

\bibitem[{\citenamefont{Guissani and Guillot}(1996)}]{GG96}
\bibinfo{author}{\bibfnamefont{Y.}~\bibnamefont{Guissani}} \bibnamefont{and}
  \bibinfo{author}{\bibfnamefont{B.}~\bibnamefont{Guillot}},
  \bibinfo{journal}{J. Chem. Phys.} \textbf{\bibinfo{volume}{104}},
  \bibinfo{pages}{7633} (\bibinfo{year}{1996}).

\bibitem[{\citenamefont{Allen and Tildesley}(1989)}]{AT89}
\bibinfo{author}{\bibfnamefont{M.}~\bibnamefont{Allen}} \bibnamefont{and}
  \bibinfo{author}{\bibfnamefont{D.}~\bibnamefont{Tildesley}},
  \emph{\bibinfo{title}{Computer Simulation of Liquids}}
  (\bibinfo{publisher}{Oxford University Press}, \bibinfo{address}{Oxford},
  \bibinfo{year}{1989}).

\bibitem[{\citenamefont{Tse et~al.}(1995)\citenamefont{Tse, Klug, and
  Allan}}]{TK95}
\bibinfo{author}{\bibfnamefont{J.~S.} \bibnamefont{Tse}},
  \bibinfo{author}{\bibfnamefont{D.~D.} \bibnamefont{Klug}}, \bibnamefont{and}
  \bibinfo{author}{\bibfnamefont{D.~C.} \bibnamefont{Allan}},
  \bibinfo{journal}{Phys. Rev. B} \textbf{\bibinfo{volume}{51}},
  \bibinfo{pages}{16392} (\bibinfo{year}{1995}).

\bibitem[{\citenamefont{Keskar and Chelikowsky}(1992)}]{KC92}
\bibinfo{author}{\bibfnamefont{N.}~\bibnamefont{Keskar}} \bibnamefont{and}
  \bibinfo{author}{\bibfnamefont{J.}~\bibnamefont{Chelikowsky}},
  \bibinfo{journal}{Phys. Rev. B.} \textbf{\bibinfo{volume}{46}},
  \bibinfo{pages}{1} (\bibinfo{year}{1992}).

\bibitem[{\citenamefont{Smyth et~al.}(1987)\citenamefont{Smyth, Smith, Artioli,
  and Kvick}}]{SSAK87}
\bibinfo{author}{\bibfnamefont{J.}~\bibnamefont{Smyth}},
  \bibinfo{author}{\bibfnamefont{J.}~\bibnamefont{Smith}},
  \bibinfo{author}{\bibfnamefont{G.}~\bibnamefont{Artioli}}, \bibnamefont{and}
  \bibinfo{author}{\bibfnamefont{A.}~\bibnamefont{Kvick}}, \bibinfo{journal}{J.
  Chem. Phys.} \textbf{\bibinfo{volume}{91}}, \bibinfo{pages}{988}
  (\bibinfo{year}{1987}).

\bibitem[{\citenamefont{Press et~al.}(1992)\citenamefont{Press, Teukolsky,
  Vetterling, and Flannery}}]{NR92}
\bibinfo{author}{\bibfnamefont{W.}~\bibnamefont{Press}},
  \bibinfo{author}{\bibfnamefont{S.}~\bibnamefont{Teukolsky}},
  \bibinfo{author}{\bibfnamefont{W.}~\bibnamefont{Vetterling}},
  \bibnamefont{and} \bibinfo{author}{\bibfnamefont{B.}~\bibnamefont{Flannery}},
  \emph{\bibinfo{title}{Numerical recipes in FORTRAN: the art of scientific
  computing}} (\bibinfo{publisher}{Cambridge University Press},
  \bibinfo{address}{Cambridge}, \bibinfo{year}{1992}).

\end{thebibliography}

\end{document}